# THE PXIE LEBT DESIGN CHOICES*

L. Prost[#], A. Shemyakin, Fermilab, Batavia, IL 60510, USA

*Abstract*

Typical front-ends of modern light-ion high-intensity accelerators typically consist of an ion source, a Low Energy Beam Transport (LEBT), a Radiofrequency Quadrupole (RFQ) and a Medium Energy Beam Transport (MEBT), which is followed by the main linac accelerating structures. Over the years, many LEBTs have been designed, constructed and operated very successfully. In this paper, we present the guiding principles and compromises that lead to the design choices of the PXIE LEBT, including the rationale for a beam line that allows un-neutralized transport over a significant portion of the LEBT whether the beam is pulsed or DC.

## INTRODUCTION

PXIE is a test accelerator under construction dedicated to the validation of various concepts envisioned for the front-end of a 2-mA, 800 MeV H$^-$ CW superconducting RF (SRF) linac proposed as part of a program of upgrades to FNAL's injection complex (a.k.a. PIP-II [1]).

The main beam requirements for the PXIE LEBT are summarized in Table 1. Its nominal mode of operation is DC. However, for commissioning purposes, the LEBT must be able to provide a wide range of duty factors, which can be adjusted by varying the pulse length and/or pulse frequency. For PIP-II, two ion sources interchangeable within minutes are required in order to increase the beam uptime. For economic reasons, only one ion source was planned for PXIE, although all the beam line features needed for accommodating a second one are included.

Table 1: PXIE LEBT parameters

| Parameter | Value | Unit |
|---|---|---|
| Ion type | H$^-$ | |
| Kinetic energy | 30 | keV |
| Nominal/Max. beam current, DC | 5/10 | mA |
| Output transverse emittance (normalized, rms) | < 0.18 | mm mrad |
| Pulse width | 0.001-16.6 | ms |
| Kicker pulse rise and fall times | ≲ 100 | nsec |
| Pressure (RFQ entrance) | ~10$^{-7}$ | Torr |

It should be noted that preliminary design studies for a LEBT of a previous incarnation of PIP-II were undertaken at LBNL [2], as part of a national collaboration, and that some of the concepts remain (e.g.: solenoid focusing, 2 ion sources).

## PXIE LEBT BEAM LINE DESIGN

*Guiding principles and first implications*

In the case of PXIE, high reliability of the beam line and its components was emphasized from the start. With that mind-set, an important aspect of the design of the PXIE LEBT was to maintain as good a vacuum as possible in the RFQ as well as to limit particle bombardment on the RFQ vanes in order to reduce the frequency of sparking. As a result, a fairly long LEBT (~2 m) was envisioned in order to isolate the inherently high vacuum pressure found near the ion source exit from the low vacuum pressure expected at the entrance of the RFQ.

Also, to limit the potential for catastrophic failures and decrease irradiation of the cavities during tuning, we think that it is highly desirable to be able to tune the SRF cryomodules using a short pulse. However, because the neutralization process is not instantaneous, the optimal tune for a short pulse may noticeably differ from long pulses or DC operation. Thus, we propose a transport scheme in which the beam is not neutralized in the downstream part of the LEBT independently of its time structure.

*Beam line layout*

A simple diagram of the beam line is pictured on Figure 1, showing its overall geometry and the placement of the main elements.

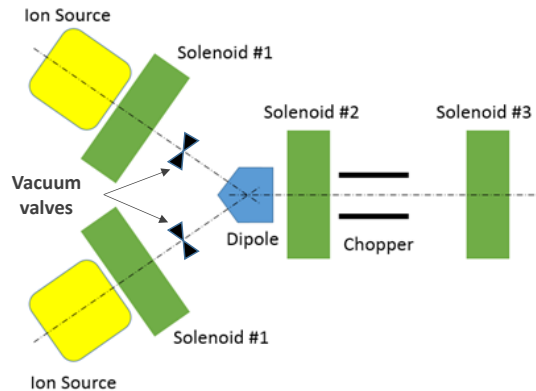

Figure 1: Conceptual schematic of the PIP-II/PXIE LEBT with 2 ion sources.

The design incorporates 2 ion sources in a "Y"-configuration with a slow switching bending dipole magnet. Each leg includes a vacuum valve that isolates the ion source from the rest of the LEBT, which allows using one source while the other is not operational.

In addition, the bend ensures that there is no direct line of sight between the ion source and the RFQ (as well as the superconducting elements further downstream). As a result, bombardment of the RFQ vanes by fast neutrals or acceleration of positive ions [3] are virtually eliminated, which should help the overall reliability of the accelerator. Also, the bending dipole can be used for personnel protection since, when turned off, the beam cannot reach the RFQ and be accelerated.

Both magnetic and electrostatic focusing lenses have been used in light-ion, low-energy, high-intensity LEBTs [4], and offer different advantages and challenges (e.g.: [5]). For the PXIE LEBT, it was decided to use solenoidal lenses. At this stage of the discussion, one of the main reasons for this choice is related to possible difficulties dealing with static high voltages near an ion source that produces a relatively high gas load. Also, one may argue that solenoids are better suited than electrostatic lenses for a long transport line. While PXIE's solenoids are otherwise pretty standard, each one incorporates a pair of dipole correctors for steering.

The primary function of the LEBT is to match the beam's optical functions between the ion source and the RFQ. From a theoretical point of view, for an axisymmetric system, only two free parameters are needed for matching purposes. Thus, one solenoid for focusing is enough, the second free parameter being the distance between the elements. However, once distances are fixed, such a setup has no flexibility; in particular, it cannot accommodate any variations of the beam parameters coming out of the ion source. Hence, in most practical applications, the minimum number of focusing lenses is two.

The need for additional focusing elements then depends on what other functions the LEBT must accommodate. We opted for a beam line with 3 solenoids (per leg) with the main argument being the flexibility of tuning it provides for transport schemes with different neutralization pattern.

Figure 2 shows the beam line as it will be implemented. For vacuum pumping, there are three 1000 l/s turbo pumps attached to the ion source vacuum chamber, and one on the chopper assembly. The vacuum pressure in the ion source is on the order of a few mTorr, down to $1\times10^{-7}$ Torr at the chopper assembly and $5\times10^{-8}$ Torr measured in the RFQ (most upstream gauge). More details about the beam line elements can be found in [6] and references therein.

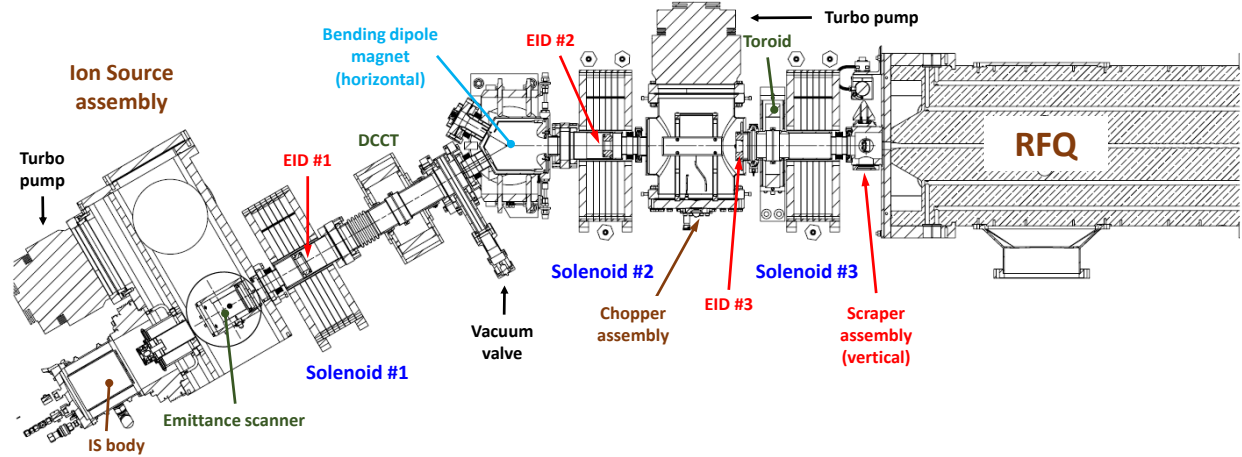

Figure 2: PXIE LEBT & RFQ (1st quarter) sectioned at the centre line.

*Chopping system*

For accelerator front-ends with similar beam parameters as PXIE's, most LEBTs employ a transport scheme that relies on almost complete neutralization of the beam to counteract the effects of the space charge during transport. However, neutralization is inevitably broken in the vicinity of a chopper. In order to decrease the distance that the beam travels with full space charge and low energy, the chopping system is often located just upstream of the RFQ. Such location has several features contradicting the principles we have set. First, absorption of the cut-out beam creates a significant gas load into the RFQ. Second, the short distance required between the last solenoid and RFQ limits the possibilities of fitting in a simple and robust chopper. It also makes it difficult to place diagnostics downstream of the chopper.

Following this logic, we placed the chopping system between solenoids #2 and #3. That allows for diagnostics to be added both before (toroid) and after (scraper) solenoid #3, as well as provides room for a vacuum pump. In recent tests, the comparison of the toroid current with an identical toroid downstream of the RFQ allowed accurate measurements of the RFQ transmission, and the scraper was found to be very useful for checking the properties of the beam injected into the RFQ.

Typically, chopping systems decouple the kicker and the absorber i.e. the beam is deflected onto a dedicated electrode (a.k.a. absorber) located downstream. The simplest design for a kicker is where two electrodes face one another and high voltage is applied across the gap. With a maximum beam power of 300 W (DC) in the PXIE LEBT, beam losses to the electrodes quickly become an issue. The solution that was adopted is to design the kicker electrodes such that one of them would also be the beam absorber. In addition to inherently making the kicker robust, it removes the need for a dedicated absorber electrode downstream, thus making the overall chopping system more compact. The absorber plate is at the ground potential but electrically isolated in order to measure the primary beam current. When the beam is passed through, the design of the chopper electronics allows applying a -300V voltage to the kicking plate to clear secondary ions out of the beam path.

### Electrically isolated diaphragms

The beam line includes 3 water-cooled, Electrically Isolated Diaphragms (EID), two of which are located within Solenoids #1 and #2.

The primary function of these electrodes is to minimize uncontrolled beam losses. They were sized such that if there is some beam loss, for instance before the optics are properly understood, or during the rise and fall time of the kicker (EID #3), it would most likely occur first at these locations and be measured.

Second, the EIDs play the role of potential barriers in the transport scheme with an un-neutralized section, confining ions in sub-sections of the beam line. Note that depending on the respective values of the applied voltages, various neutralization patterns can be obtained, allowing to explore and compare various transport schemes, including one where the beam would be neutralized over the entire beam line.

Finally, they are used to measure the beam size and centre the beam. This is achieved by steering the beam with upstream correctors and recording the current drawn by the electrodes.

In addition, there is a so-called scraper assembly just downstream of Solenoid #3, which is an additional movable electrically-isolated and water-cooled electrode with 3 apertures. It enables estimating and controlling the beam size at the RFQ entrance.

### Emittance growth mitigation

To satisfy both the beam physics requirements and the design choices for PXIE, a hybrid transport scheme has been devised, in which the beam propagates through the first, 'high pressure' part of the LEBT being neutralized, but neutralization is stopped right upstream of the chopper, and further propagates in the 'low pressure' part of the LEBT completely un-neutralized.

In reference [7], it is argued that for an ion source generating a beam with a uniform current density and Gaussian velocity distribution, a completely neutralized beam transport with linear optics replicates the uniform current density distribution of the beam in the image plane of the first solenoid. If the beam perveance is modest, neutralization can be removed in this location and the beam will propagate with negligible emittance growth over some distance only determined by the spread of the velocity distribution.

Note that the PXIE LEBT permits neutralized transport over its entire length as well.

### Machine protection

The LEBT chopper is used as the main device to interrupt the beam when the Machine Protection System (MPS) issues a fault signal. Because it is not fail-safe (the beam passes through if its power supply is broken or cable disconnected), we plan to implement a two-tier MPS, in which the fault signal also commands to switch off the ion source extraction electrode power supply. In case no interruption is detected, the ion source bias and bend power supplies are also turned off.

## CONCLUSION

The choices made during the design of the PXIE LEBT were guided by the goal of making a reliable, tuneable, and flexible machine. Its atypical layout with the possibility of accommodating two ion sources, a chopper located ~1m away from the RFQ, and 3 EIDs in the beam line relies on a partially un-neutralized transport scheme in order to generate a beam with constant beam parameters independent on the time structure.

The LEBT has been assembled and fully commissioned in a straight configuration (without the bend yet). Measurements presented in [8] support that this scheme works with the PXIE LEBT as designed.

## ACKNOWLEDGMENT

Authors acknowledge the preliminary design work carried out by J. Staples, Q. Ji and D. Li from LBNL, and the subsequent discussions and critiques that were essential for the successful realization of the design presented here. We are grateful to R. Andrews, D. Snee, T. Hamerla and their teams for the mechanical designs of many elements, their fabrication and installation.

## REFERENCES

[1] PIP-II Reference Design report, PIP-II Document database, Doc. #1
[2] Project X Collaboration Meeting, 9-10 September, 2010, Fermilab (2010) *several contributions*
[3] M. Plum, IPAC2013, Shanghai, China, (2013) MOXBB101
[4] L. R. Prost, FERMILAB-TM-2622-AD (2016), arXiv:1602.05488 [physics.acc-ph]
[5] Chauvin et al., LINAC2010, Tsukuba, Japan (2010) TH302
[6] L. Prost et al., IPAC'14, Dresden, Germany (2014) MOPRI086
[7] A. Shemyakin, L. Prost, FERMILAB-TM-2599-AD (2015), arXiv:1504.06302 [physics.acc-ph]
[8] L. Prost, J.-P. Carneiro, A. Shemyakin, *these proceedings*, TUPMR033